\documentclass{emulateapj}
\usepackage{graphicx}

\newcommand{\beq}	{\begin{equation}}
\newcommand{\eeq}	{\end{equation}}
\newcommand{\beqa}{\begin{eqnarray}}
\newcommand{\eeqa}{\end{eqnarray}}
\newcommand{\avg}[1]  {{\langle #1 \rangle}} 

\newcommand{\e}	{$^{-1}$}
\newcommand{\ee}	{$^{-2}$}
\newcommand{\eee}	{$^{-3}$}
\def\simlt{\lower.5ex\hbox{$\; \buildrel < \over \sim \;$}}
\def\simgt{\lower.5ex\hbox{$\; \buildrel > \over \sim \;$}}
\def\la{\simlt}

\font\tenbi=cmmib10 
\newfam\bifam  \textfont\bifam=\tenbi

\font\tenbr=cmbx10
\newfam\brfam  \textfont\brfam=\tenbr

\font\squinttenbi=cmbx10 at 9pt
\scriptfont\brfam=\squinttenbi

\def\vecnabla{
             \setbox1=\hbox{$\bigtriangledown$}
                          \raise.45ex\hbox{$\bigtriangledown$\hskip-.97\wd1
                          $\bigtriangledown$\hskip-.97\wd1
                          $\bigtriangledown$\hskip-.97\wd1}
                          \raise.47ex\hbox{$\bigtriangledown$}}

\def\msun	{M_\odot}
\def\lsun	{L_\odot}

\newcommand{\caln}		{{\cal N}}
\newcommand{\calr}		{{\cal R}}
\newcommand{\pbyp}[1]	{{{\partial\hfil}\over{\partial#1}}}
\newcommand{\ppbyp}[2]	{{{\partial#1}\over{\partial#2}}}

\newcommand{\crdm}		{{\cal R}_{\dot m}}
\newcommand{\ecore}	{\epsilon_{\rm core}}
\newcommand{\jf}		{{j_f}}
\newcommand{\krho}		{{k_\rho}}
\newcommand{\mf}		{{m_f}}
\newcommand{\ml}		{{m_\ell}}
\newcommand{\mfl}		{{m_{f,\ell}}}

\newcommand{\mmax}	{{m_{p,\,\rm max}}}
\newcommand{\mup}		{{m_u}}

\newcommand{\mdo}		{\dot m_1}
\newcommand{\mds}		{\dot m_{\rm IS}}
\newcommand{\mdtc}	{\dot m_{\rm TC}}
\newcommand{\nds}		{\dot\caln_*}
\newcommand{\nh}		{n_{\rm H}}
\newcommand{\nbhf}		{\bar n_{\rm H,\, 4}}

\newcommand{\phipc}	   {\phi_{P,\,\rm core}}
\newcommand{\ppt}		{\psi_{p2}}

\newcommand{\psal}		{\psi_{\rm Sal}}
\newcommand{\rmd}		{\calr_{{\dot m}}}
\newcommand{\scl}		{\Sigma_{\rm cl}}
\newcommand{\tacc}		{t_{\rm acc}}
\newcommand{\tff}		{t_{\rm ff}}
\newcommand{\tfo}		{t_{f1}}
\newcommand{\tf}		{t_f}
\newcommand{\ndso}		{\dot\caln_{*,0}}
\newcommand{\obs}		{{\rm obs}}
\newcommand{\ii}			{{\rm II}}

\shorttitle{The Protostellar Mass Function}
\shortauthors{McKee \& Offner}
\begin{document}

\title{The Protostellar Mass Function}
\author{Christopher F. McKee}
\affil{Physics Department and Astronomy Department, University of California,
   Berkeley, CA 94720}
\email{cmckee@astro.berkeley.edu}
\and
\author{Stella S. R. Offner}
\affil{Harvard-Smithsonian Center for Astrophysics, Cambridge, MA 02138 }
\email{soffner@cfa.harvard.edu }

\begin{abstract}

The protostellar mass function (PMF) is the Present-Day Mass Function of
the protostars in a region of star formation. It is determined by the initial mass
function weighted by the accretion time. The PMF thus depends on the
accretion history of protostars and in principle provides a powerful tool
for observationally distinguishing
different protostellar accretion models. We consider three 
basic
 models here:
the Isothermal Sphere model \citep{shu77}, the Turbulent Core model 
\citep{mck03}, and 
an approximate representation of
the Competitive Accretion model \citep{bon97, bon01a}.
We also consider modified versions of these accretion models, in which  
the accretion rate tapers off linearly in time. Finally, we allow for an overall
acceleration in the rate of star formation. At present, it is not possible to directly
determine the PMF since protostellar masses are not currently measurable. 
We carry out an approximate comparison of predicted PMFs
with observation by using the theory to infer the conditions in the ambient
medium in several star-forming regions. 
Tapered and accelerating models generally agree better with observed
star-formation times than models without tapering or acceleration, but
uncertainties in
the accretion models and in the observations do not allow one to 
rule out any of the proposed models
at present.
The PMF is essential for the calculation
of the Protostellar Luminosity Function, however, and this enables stronger
conclusions to be drawn \citep{off10}.

\end{abstract}

\section{Introduction}

The Initial Mass Function (IMF) is of central importance in star formation since the
properties of a star and its effects on the surrounding medium
are determined primarily by its initial mass. One of the main approaches for inferring
the IMF is to apply evolutionary models to observations of
the Present-Day Mass Function (PDMF) of a group of
stars. The IMF is created during the process of star formation, when the mass of each
protostar grows by accretion until it reaches its final value.
Observations of a region of star formation can, in principle, allow
one to infer the PDMF of the protostars there; we refer to this as
the Protostellar Mass Function (PMF). The PMF depends on both the
IMF, which determines the relative number of stars when the star formation
is complete, and on the process of star formation, which determines how long
each protostar spends at a given mass. Because of this latter property,
the PMF is potentially a powerful tool for inferring the nature of the 
star formation process. For example, inside-out collapse of
an isothermal sphere \citep{shu77} leads to protostellar lifetimes that
are proportional to the mass of the final star, whereas models based
on competitive accretion \citep{zin82,bon97} have protostellar lifetimes
that are independent of the mass of the final star;
the inside-out collapse of a turbulent core \citep{mck02,mck03} has
an intermediate behavior.
As we shall see
below, it follows that the PMFs predicted by these models are quite different.

There are of course other approaches for observationally distinguishing among
the different theories of star formation. One is to study the relation between
the mass distribution of density concentrations in molecular clouds
(the Core Mass Function, or CMF) and the stellar IMF, which are observed to be similar
\citep{mck07}. This similarity is the basis for recent theories of the IMF,
which are predicated upon the assumption that stellar masses are determined
by the production of gravitationally bound cores in turbulent molecular
clouds \citep{pad02,pad07,hen08,hen09}. Such a direct connection
between the CMF and IMF would appear to be a natural prediction
of theories based on the inside-out collapse of gravitationally bound
cores \citep{shu77,mck02}, but inconsistent with the theory of competitive
accretion (e.g., \citealp{bon97}).  Indeed, in inside-out collapse
theories, the PMF and CMF are closely related, since protostars are forming
in a significant fraction of cores. \citet{cla07} have pointed out that the 
mass distribution of cores depends on their lifetime, just as we shall see
below that the PMF depends on the accretion time, and that this dependence
would make the slope of the IMF significantly steeper than that of the CMF.
However, in the Turbulent Core model
\citep{mck03} the star formation time depends only weakly on the mass
of the core, and \citep{hen09} have argued that as a result the
slopes of the CMF and IMF would agree within the observational errors.
It is clear that additional approaches for testing theories of star
formation are needed.

The principal difficulty with the PMF method developed here is that
at present, it is difficult to infer the masses of individual
protostars, both because the evolutionary tracks of protostars are uncertain
(e.g., \citealp{cha07}) and because of the effects of accretion on the spectrum
of the protostar, which are difficult to quantify.
These difficulties will presumably be overcome in the future. Currently,
the best way to determine the PMF appears to be through observations
of the Protostellar Luminosity Function (PLF), which wlll be discussed in
Offner \& McKee (2010; hereafter Paper II). 

In this work, we define a protostar as an embedded source that is
still experiencing signficant accretion and thus is more than a few
percent from its initial stellar mass. Observationally, young stellar
objects are characterized on the basis of the slopes of their spectral
energy distributions into four classes, 0-III \citep{ada87,andre94}. However, since geometric
effects due to disk and outflow orientation influence the radiation
reprocessing of the embedded source, it is difficult to directly map
classes to physical stages. For the latter, we use the classification 
proposed by \citet{cra08}: 
Stage 0
objects have protostellar masses that are less than or equal to their
envelope, while Stage I objects have protostellar masses larger than
the envelope. Once the envelope mass falls below 0.1
$\msun$, the object is considered to have completed its main
accretion phase and entered the Stage II phase. Although these stages do not
directly correlate with class definitions, in practice, Class 0 and Class I
sources approximately correspond to Stage 0 and Stage I (see \S 6
for additional discussion).

In our analysis, we shall make 
two main
assumptions: First, we shall generally
adopt a \citet{cha05} IMF with an imposed upper mass cutoff at $\mup$.
This IMF has a log-normal form below $1 M_\odot$ and a power-law form above.
For star formation regions large enough to fully sample the IMF,
$\mup\sim 150 M_\odot$ \citep{fig05}. However, 
as we show below,
regions of low-mass star
formation often have maximum stellar masses of only a few $M_\odot$,
even though in some cases one would expect stars more massive than that according
to the \citet{cha05} IMF.
It does not appear that this is a selection effect, since the data we compare with
\citep{eva09} represent a complete survey of nearby star-forming regions. 
If the deficit of high-mass stars is not a selection effect or a rare statistical
fluctuation, then it must be
the result of a physical
inability to form high-mass stars in some regions---e.g., \citet{kru08}.
In any case, our assumption is that the IMF above $1 M_\odot$ in these regions can be approximated by a truncated power-law
im which the upper limit on the final masses of the protostars is set by
the inferred upper limit on the masses of the more numerous 
newly formed stars (primarily Class II sources)
in the sample.
In the applications to observations below, we shall generally adopt an upper mass
limit of $3 M_\odot$.
Second, we shall assume that
the accretion rate onto the protostar can be expressed as a simple function of
the instantaneous protostellar mass, the final protostellar mass (i.e., the
initial stellar mass), and the time. 
In particular, 
we ignore complications
associated with an initial Larson-Penston accretion phase, when the accretion
rate can be much larger than the value expected on the basis of dimensional
analysis (e.g., \citealp{mck07};
however, in one of the most
complete simulations of the formation of a star to date, \citet{mac09} found
only a small enhancement in the accretion rate at  early times);
Furthermore, we average over any temporal fluctuations in the accretion rate, such
as occur in FU Ori outbursts \citep{har96}.
In principle, the accretion rate can depend on the location of the protostar in
its natal cluster; in our treatment, any such dependence is encoded in the
dependence on the final mass of the protostar.

The protostellar mass and luminosity functions were first considered
by \citet{fle94a,fle94b}.
In their two companion papers, they derived the time-dependent mass and luminosity
functions for young embedded stellar clusters, including the luminosity
contributions from protostars through
main-sequence stars.
Our work differs from theirs in several important respects:
(1) We consider a variety of different theories for star formation, 
allowing for the accretion rate to depend on the protostellar mass, $m$, and the
final stellar mass, $m_f$, whereas they
assumed that the accretion rate for all the stars was
constant. 
Non-constant accretion 
allows us to
test different theories of star formation. In particular, as noted by
\citet{shu87}, the constant accretion-rate model, which was developed
for low-mass star formation, is unlikely to apply to high-mass star
formation. 
(2) As we mention
above, observations in
the solar neighborhood suggest an upper cutoff in the mass distribution at a few
solar masses; we explicitly allow for this in our adopted IMF, whereas
Fletcher \& Stahler did not. 
(3) They treated the protostellar,
pre-main sequence
and main sequence stages, 
whereas we consider only the protostellar case. 
(4) They focused on the case in
which the star formation rate is constant for a given time period, whereas we consider
both the steady-state case and the case of accelerating star formation
\citep{pal99,pal00}. 
(5) Finally, we note a difference in approach: they determined
the probability that a star of a given mass would cease accreting and become
a pre-main sequence star, whereas we label each protostar by its instantaneous mass
and its final mass. As we shall see below, this makes it straightforward for us
to consider the case of tapered accretion, in which the accretion rate
declines prior to the end of the protostellar stage.

We begin with the derivation of the PMF in terms of the IMF and the accretion
history of the protostars in \S 2. We consider 
four accretion histories in
\S 3: the collapse of an isothermal sphere \citep{shu77}, the Turbulent
Core model \citep{mck02,mck03}, a blend of these two
models, 
and the Competitive Accretion model \citep{bon97}.
We also consider the effects of tapering the accretion rate as
the mass approaches the final mass, as suggested by \citet{mye98}.
In \S 4, we evaluate the PMF for these accretion histories, both analytically
for a \cite{sal55} IMF and numerically for a \citet{cha05} IMF. \citet{pal00} have
suggested that nearby low-mass star-forming regions have star formation rates
that are accelerating in time, and we show how this affects the PMF in \S 5.
We make a brief comparison with observation in \S 6 (a more extensive comparison
will be given in Paper II) and summarize our conclusions in \S 7.

\section{Steady-State Protostellar Mass Function}

Consider a region in which stars are forming. We shall refer to this group
of stars as a cluster, although we make no assumption as to whether the group of
stars is gravitationally bound.
In this section, we assume that star formation is in a steady state,
so that there are both stars that have reached
their final mass and protostars. 
For clarity, we list our mathematical symbols and their
definitions in Table \ref{table0}. Each
protostar begins with a negligible initial mass and accretes until it reaches a final mass
$m_f$. The cluster IMF describes the distribution of stars in the cluster
with respect to their final mass:
in a range of final masses $d\mf$, the rate at which stars are forming is
\beq
d\dot\caln_*=\dot\caln_*\psi(m_f) d\ln m_f,
\eeq
where $\dot\caln_*$ is the total star formation rate (i.e., the number of stars
forming per unit time). In the cluster,
stellar masses extend from a lower limit $\ml$
(theoretically expected to be $\sim 0.004 M_\odot$---\citealp{low76} 
as updated by \citealp{whi07})
to an upper limit $\mup$, which we infer from observations of the cluster.
Note that the IMF is normalized to unity,
\beq
\int_\ml^\mup \psi(m_f) d\ln\mf =1.
\eeq

\begin{figure}
\epsscale{1.20}
\plotone{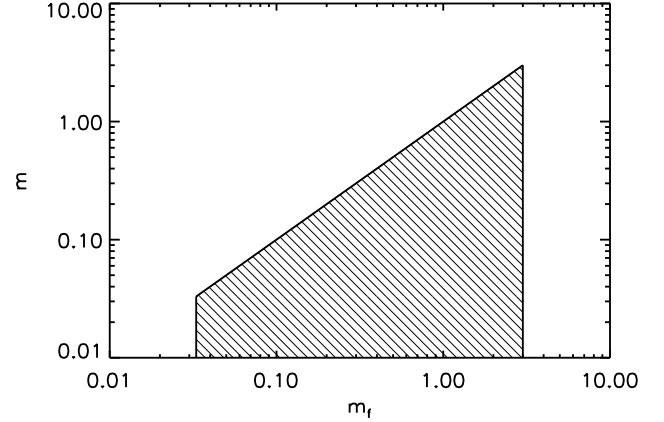}
\caption{ \label{mvsm} The domain of integration for the PMF (eqs. \ref{eq:pmfnorm1}, 
\ref{eq:mmf}),  At each value of
the final protostellar mass, $\mf$, integration over $m$ extends from $m=0$ to
$m=\mf$. For each value of the current mass of the protostar, $m$, the
integration over $\mf$ extends from $\mfl=\max(m,\ml)$ to 
the maximum mass of stars in the cluster,
$\mup$.
In the plot, we adopt $\ml = 0.033$ and $\mup=3$.}
\end{figure}

	As discussed in \S 1, one of our assumptions is that, in a steady
state, the accretion history of 
a protostar is determined by two parameters, its mass and its final mass.
The distribution of protostars is therefore completely described by the
bivariate PMF, $\ppt(m,\mf)$, such that the number of protostars in
the mass range $d m$ with final masses in $d\mf$ is
\beq
d^2\caln_p=\caln_p\ppt(m,\mf) d\ln m\; d\ln \mf,
\label{eq:dnp1}
\eeq
where $\caln_p$ is the total number of protostars in the cluster.
The normalization of the bivariate PMF can be expressed in two forms:
\beqa
\int_\ml^\mup d\ln\mf \int_0^\mf d\ln m \;\ppt(m,\mf)&=&1, 
\label{eq:pmfnorm1}\\
\int_0^\mup d\ln m\int_\mfl^\mup d\ln\mf\; \ppt(m,\mf)&=&1,
\label{eq:mmf}
\eeqa
where the ranges of integration explicitly impose the condition that $\mf \geq m$
and where 
\beq
\mfl\equiv \max(m,\ml)
\label{eq:mfl}
\eeq
is the lower bound on the integration over $\mf$ 
in equation (\ref{eq:mmf}).
The region of integration in the $m-\mf$ plane is shown in Figure \ref{mvsm}.

	The PMF is the present day mass function of the protostars in
the cluster,
\beq
\psi_p(m)=\int_\mfl^\mup \ppt(m,\mf) d\ln \mf;
\label{eq:pmf1}
\eeq
it is normalized to unity also, as ensured by equation (\ref{eq:mmf}).
The PMF is observable in principle, although that is currently difficult as
discussed in \S 1. 

	To determine the bivariate PMF, $\ppt$, note that the number of
protostars born in a time interval $dt$ that will have final masses
in the range $d\mf$ is
\beq
d^2\caln_p=\nds\psi(\mf)dt\; d\ln\mf.
\label{eq:dnp2}
\eeq
If we let $t_f(\mf)$ be the time required to form a star of mass $\mf$, then
integration of this equation gives the total number of protostars as
\beqa
\caln_p &=&\nds\int_\ml^\mup d\ln\mf\psi(\mf) t_f(\mf),\\
&\equiv & \nds\avg{t_f},
\label{eq:np}
\eeqa
where we denote the average of some quantity $x$ over the IMF as $\avg{x}$.
Now, the characteristic accretion time scale for a protostar of
mass $m$ and final mass $\mf$ is
\beq
\tacc(m,\mf)\equiv \frac{m}{dm/dt}=\frac{dt}{d\ln m},
\label{eq:tacc}
\eeq
so that equation (\ref{eq:dnp2}) becomes
\beq
d^2\caln_p=\nds\psi(\mf)\tacc\; d\ln m\; d\ln\mf.
\label{eq:d2n}
\eeq
Equations (\ref{eq:dnp1}), (\ref{eq:dnp2}) and (\ref{eq:d2n}) then give the final expression
for the bivariate PMF,
\beq
\ppt(m,\mf)=\frac{\psi(\mf)\tacc(m,\mf)}{\avg{t_f}}.
\label{eq:ppt}
\eeq
This expression can be readily generalized to the case in which the accretion rate
depends on more than two variables.

The PMF (eq. \ref{eq:pmf1}) is then
\beq
\psi_p(m)=\frac{1}{\avg{t_f}}\int_\mfl^\mup\psi(\mf)\tacc(m,\mf)\; d\ln\mf,
\label{eq:pmf2}
\eeq
where $\mfl$ is defined in equation (\ref{eq:mfl}).
{\it The PMF is thus the IMF weighted by the accretion time, $\tacc=m/\dot m$,
for all stars with final masses exceeding $m$}.
Note that as $m\rightarrow\mup$ the value of the integral approaches
zero: there are very few protostars with masses close to the maximum value.

\begin{deluxetable*}{cl}
\tablecolumns{2}
\tablecaption{Symbol Key \label{table0}}
\tablehead{ \colhead{Symbol} & \colhead{Definition} }\\
\startdata

$\psi_C$ 		& Chabrier (2005) initial mass function \\
$m$ & Instantaneous protostellar mass of a given star ($\msun$) \\
$\mf$ &  Final protostellar mass = initial stellar mass of a particular star ($\msun$)
\\
$\ml$ & Lowest observable stellar mass ($\msun$)\\
$\mup$ & Highest mass star that can be formed in the cluster ($\msun$)\\
$\caln_p$ & Number of protostars
\\
$\nds$  & Star formation rate (number per unit time)\\
$\tf$   & Time to form a star of mass $\mf$ (yr) \\
$\tfo$  & Time to form a star of 1 solar mass without tapering (yr) \\
$\dot m$ & Instantaneous protostellar accretion rate ($M_\odot$ yr$^{-1}$)\\
$\mdo$  & Final accretion rate of a 1 solar mass star without tapering ($M_\odot$ yr$^{-1}$) \\
$\ppt$ & Bivariate protostellar mass function in terms of $m$ and $\mf$ \\
$\psi_p$ & Protostellar mass function in terms of $m$ \\

\enddata
\end{deluxetable*}

\section{Accretion Histories}
\label{sec:accretion}

\subsection{Power-Law Accretion}
\label{sec:power}

The PMF depends on both the accretion time, $\tacc$, and on the mean
formation time, $\avg{t_f}$ (eq. \ref{eq:pmf2}), so 
determining it requires evaluating the accretion histories of the protostars in the cluster.
Several different models for protostellar accretion have been proposed,
and indeed measurement of the PMF would provide a powerful method
for distinguishing among them.

The most commonly used model for low-mass star formation is
the inside-out collapse 
of a singular isothermal sphere \citep{shu77};
we term this the Isothermal Sphere model.
In this
model, gas accretes from an isothermal gas in 
hydrostatic equilibrium (a ``core") onto the protostar at a constant rate determined
by the temperature of the medium,
\beq
\dot m= \mds=1.54\times 10^{-6} T_1^{3/2}~~~M_\odot~\mbox{yr\e},
\label{eq:mds}
\eeq
where $T_1\equiv T/(10~$K). This expression is based on the assumption
that the gas is initially static. \citet{hun77} generalized the Shu solution
to times prior to the initial formation of the protostar, so that
at the time of protostar formation the gas is in subsonic collapse.
The most rapid collapse, at about 1/3 the sound speed, has an accretion
rate 2.6 times greater than the Shu value. 
Furthermore, \citet{li97} suggest that magnetic fields increase the accretion rate for low-mass protostars, typically by about a factor 2.
On the other hand,
protostellar outflows are likely to eject some of the core material,
reducing the final protostellar mass to a fraction $\ecore
M_{\rm core}$, where $\ecore$ is the core star formation efficiency.
\citet{mat00} estimated theoretically that $\ecore\sim 0.25-0.75$;
in a recent detailed study of cores in the Pipe Nebula, \citet{rat09}
infer $\ecore=0.22\pm 0.08$ there. The effects of initial infall and
magnetic fields on the one hand and
protostellar outflows on the other
tend to cancel, but they render the estimate
of the accretion rate in equation (\ref{eq:mds}) somewhat uncertain.

Since the time required to form a star via isothermal accretion 
scales directly as the mass of the star, this model does not work
for high-mass star formation \citep{shu87}.
In order to treat high-mass star formation,
McKee \& Tan (2002, 2003) developed the Turbulent Core model
as a generalization of the Isothermal Sphere model. They 
considered a gravitationally bound clump of gas in which a cluster of stars is forming.
Within this clump, individual stars form from bound cores. Both
the clump and the embedded cores were assumed to be in 
approximate virial equilibrium, supported by internal turbulent motions.
In terms of the surface density
of the clump, $\scl=M_{\rm cl}/\pi R_{\rm cl}^2$, they 
found that the typical protostellar accretion rate is 
\beq
\dot m= \mdtc \left(\frac{m}{\mf}\right)^j \mf^{3/4},
\eeq
where the coefficient
$\mdtc$ is proportional to the 3/8 power of the pressure in the star-forming core
and where
the exponent $j$ is related to the density profile of the 
core ($\rho\propto r^{-\krho}$) by
\beq
j=\frac{3(2-\krho)}{2(3-\krho)}.
\eeq
To determine $\mdtc$, they
defined $\phipc$ as the ratio of the typical pressure of a star-forming
core to the mean pressure of the clump. Since the pressure in a
self-gravitating clump varies as $\scl^2$, their result corresponds to
$\mdtc\propto\phipc^{3/8}\scl^{3/4}$.
They focused on high-mass star formation and allowed for
the observed mass segregation of such stars by assuming that 
these stars formed on average
at about 30\% of the half-mass radius of the clump; for this case, they estimated
$\phipc\simeq 2$. In the absence of such mass segregation, $\phipc\simeq 1$,
and we adopt that value here. This reduces the accretion rate from
their value by a modest amount [$(1/2)^{3/8}=0.77$]. 
With this correction, the coefficient $\mdtc$ becomes
\beq
\mdtc=2.8\times 10^{-5}\scl^{3/4}~~~M_\odot~\mbox{yr\e}.
\label{eq:mdtc}
\eeq
\citet{mck03} set the remaining parameter in the accretion rate, $j$, by fixing 
the density
power law at $\krho=\frac 32$ based on observations of clumps in which
high-mass stars are forming.
The precise value of $\krho$ is not important, however, since one can show that
the value of $\mdtc$  is within 10\% of
the value quoted in equation (\ref{eq:mdtc}) over the range $1.3\leq\krho\leq 2$.

An assumption in the Turbulent Core  model is that the turbulence
is supersonic. \citet{mck03} gave an approximate generalization
of the accretion rate
that includes the case in which the turbulence is subsonic and the accretion
approaches the Isothermal Sphere value:
\beq
\dot m=\mds\left[1+\rmd^2\left(\frac{m}{\mf}\right)^{2j}\mf^{3/2}\right]^{1/2},
\label{eq:mdtctc}
\eeq
where
\beq
\rmd\equiv\frac{\mdtc}{\mds}.
\eeq
For $j=\frac 34$, corresponding to $\krho=1$, this is similar to the TNT model of
\citet{mye92}.

A third model of accretion is Competitive Accretion, in which a group
of stars in a common gravitational potential accrete gas until it is exhausted
or ejected \citep{bon97,bon01a,bon01b,cla07,cla09}. These authors emphasize that the
accretion rate depends on the location of a protostar within the
clump of gas and on the time evolution of the density in the clump.
We take these effects into account indirectly by having the accretion rate depend upon
the final stellar mass so as to produce the prescribed IMF.
Initially the accretion rate is governed by tidal effects, so
that $\dot m\propto m^{2/3}$; some of the stars fall to the center where they
are virialized and then accrete at the Bondi-Hoyle rate \citep{bon01a},
although the central, most massive star continues to accrete at the tidal
rate \citep{cla09}. We cannot take
this change in the accretion rate into account in our model, but we note that once the
protostars are virialized, the mass accreted per dynamical time is 
small \citep{bon01a,kru05}.
An important feature of Competitive Accretion is that all the protostars cease accreting at
the about the same time, presumably due to 
stellar feedback, as explicitly stated by \citet{bon01b} in their discussion
of the IMF produced by Competitive Accretion. 
(Of course, in reality the gas removal is not instantaneous and the accretion
will turn off more gradually; this can be treated by tapering the accretion
rate, as discussed below.)
Competitive Accretion is thus a {\it constant time} accretion model; this is in
contrast to the Isothermal Sphere model, a {\it constant rate} accretion model in 
which the time for a star to form scales linearly with its mass.
In order for all stars to form in the same time, the accretion rate must satisfy
$\dot m\propto \mf$. 
The resulting model for Competitive Accretion has an accretion rate
\beq
\dot m_{\rm CA} =\mdo\left(\frac{m}{\mf}\right)^{2/3}\mf,
\label{eq:mdca}
\eeq
where $\mdo$ is the final accretion rate for  star of unit mass.
\citet{bon01a} show that the
star-formation time, $t_f$, is about equal to the initial free-fall time of the natal cloud,
\beq
t_f\simeq \tff=0.435\,\nbhf^{-1/2}~~~\mbox{Myr},
\eeq
where $\nbhf\equiv \bar\nh/(10^4$~cm\eee) and $\bar\nh$ is the mean 
density of hydrogen nuclei in the cloud.
Integration of equation (\ref{eq:mdca}) shows that the characteristic accretion rate
is related to $t_f$ by $\mdo=3/t_f$.
The principal approximation in our treatment of Competitive Accretion is
the assumption that the star formation is in a steady state or is
accelerating (\S \ref{sec:accel}). Our model thus  applies to
a cluster consisting of a number of small sub-clusters, each of which
forms competitively, or to a sample of clusters.

\begin{deluxetable*}{lccccc}
\tablecolumns{6}
\tablecaption{Accretion Models \label{table1}}\footnote{Here $\Sigma_{\rm cl,-1}\equiv\Sigma_{\rm cl}/$(0.1 g cm\ee) and $\nbhf\equiv\bar\nh/(10^4$ cm\eee). Accretion rates and time scales are for untapered accretion.}
\tablehead{ \colhead{Model} & \colhead{$~j$} & \colhead{$~j_f$} &\colhead{$\mdo$~~ ($\msun$ yr$^{-1}$)} &  \colhead{$\tfo$ ~~(Myr)} & \colhead{$\avg{\tf}$~~(Myr)}}\\
\startdata
Isothermal Sphere 		& 0	 &  0    &	$1.54 \times 10^{-6}T_1^{3/2}$	  &  0.65 $T_1^{-3/2}$	 & 0.25 $T_1^{-3/2}$ \\
Turbulent Core	 	& 1/2 &  3/4 & 	$4.9 \times 10^{-6} \Sigma_{\rm cl, -1}^{3/4}$	 
&   0.40 $\Sigma_{\rm cl, -1}^{-3/4}$ &	 0.29 $\Sigma_{\rm cl, -1}^{-3/4}$ 	 \\ 
Competitive Accretion & 2/3&  1  & 	$6.9\times 10^{-6}\nbhf^{1/2}$	 &   $\tff=0.435
\nbhf^{-1/2}$  & $\avg{t_f}=\tff$  \\ 
\enddata
\end{deluxetable*}

The
Isothermal Sphere, Turbulent Core and Competitive Accretion
models all fit the form
\beq
\dot m=\mdo\left(\frac{m}{\mf}\right)^j\mf^\jf
\label{eq:mdot}
\eeq
(see Table 2). For $j<1$, this can be integrated to give
\beq
m^{1-j}=(1-j)\mdo\mf^{\jf-j}\, t. 
\label{eq:mass}
\eeq
The time to form the star (i.e., for $m$ to reach $\mf$) is 
\beq
t_f=\tfo\mf^{1-\jf}, 
\label{eq:tf}
\eeq
where
\beq
\tfo=\frac{1}{(1-j)\mdo}
\eeq
is the time to form a $1 \, M_\odot$ star.
Note that if $\mdo$ is expressed in units of $M_\odot$~yr\e, for example,
then $\tfo$ is in units of yr. The time scales for these
models are included in Table 2,
both the time to form a $1 M_\odot$ star, $t_f(1)$, and the IMF-averaged formation time,
$\avg{t_f}$. (These time scales are for untapered accretion---see \S \ref{sec:taper} below.)
For the two-component Turbulent Core model
(eq. \ref{eq:mdtctc}), the star-formation time is
\beq
t_f=\left[\frac{2}{(1+\rmd^2\mf^{3/2})^{1/2}+1}\right]\frac{\mf}{\mds},
\label{eq:tftc}
\eeq
which approaches the Isothermal Sphere value at low masses and the Turbulent Core
value at high masses.
The accretion time is then
\beq
\tacc\equiv\frac{m}{\dot m}=(1-j)\left(\frac{m}{m_f}\right)^{1-j} t_f.
\label{eq:tacc1}
\eeq
For protostellar masses significantly below the final stellar mass, the
accretion time is longest for the Competitive Accretion model 
($j=\frac 23$) and shortest
for the Isothermal Sphere model ($j=0$).

The models we consider here are by no means exhaustive.
It is also possible to contruct entirely different models for the accretion history,
for example, those in which $j>1$, so that
the protostellar masses diverge at a finite time (e.g., \citealp{beh01});
such models require one to specify an initial protostellar mass,
which is not well-defined.
\citet{mye09} has proposed a model that synthesizes the Isothermal
Sphere and Competitive Accretion models. In this model, the protostar 
initially experiences freefall collapse where 
the accretion rate is comparable to $\dot m_S$, while at late times
the accretion rate approaches 
$\dot m_* \propto m_*^{5/3}$, which is close to the $m_*^2$ dependence
of Bondi-Hoyle accretion. (Note that in our modeling
of Competitive Accretion, we have adopted the $m_*^{2/3}$ dependence
that \citealp{bon01a} found in their numerical simulations.) 
Models such as these that
lead to an explosive growth in the stellar mass require an 
abrupt termination of the accretion.

\subsection{Tapered Accretion}
\label{sec:taper}

The accretion models presented above share one unphysical
characteristic: the accretion drops to zero discontinuously when
the protostar reaches its final mass. 
Models such as these that
lead to an explosive growth in the stellar mass require an 
abrupt termination of the accretion.
\citet{mye98}
addressed this problem by assuming that the accretion rate declined
exponentially with time, $\dot m\propto \exp-(t/t_d)$. A difficulty with
this model is that the accretion continues indefinitely. More
recently, \citet{mye08} has included a model for a time-dependent
dispersal of the protostellar core by protostellar
outflows. If the dispersal time is sufficiently short,
the protostellar mass
converges to a well-defined value.

Even in the absence of prototellar outflows, one expects the
accretion rate to taper off continuously. 
Observationally, \citet{fed09} find that accretion falls below 
$10^{-11} \msun$yr$^{-1}$ for nearly all stars by 10 Myr. 
However,  the stars gain a negligible 
amount of mass via accretion after only a couple of Myr.
For expansion wave solutions of the type in the Shu solution and
the Turbulent Core model, the self-similarity of the accretion
is broken when the expansion wave reaches the final mass, $\mf$.
If the core is immersed in a medium of uniform pressure, the
expansion wave will be reflected as a compression wave,
and the accretion rate will decrease when that wave reaches
the protostar
\citep{sta80,mcl97}. 

A simple way to incorporate the decrease in accretion due to
dispersal and boundary effects is to reduce the accretion
rate by a tapering factor $[1-(t/t_f)^n]$:
\beq
\dot m=\mdo\left(\frac{m}{\mf}\right)^j\mf^\jf\left[1-\left(\frac{t}{t_f}\right)^n\right],
\eeq
where $n>0$ and
where $\mdo$ is the final accretion rate for a $1\, M_\odot$ star in the absence of
tapering.
Integration of this relation gives
\beq
m^{1-j}=(1-j)\mdo\mf^{\jf-j}\left[1-\frac{1}{n+1}\left(\frac{t}{t_f}\right)^n\right]t.
\eeq
The protostar reaches its final mass at a time
\beqa
t_f&=&\left(\frac{n+1}{n}\right)\frac{\mf^{1-\jf}}{(1-j)\mdo},\\
&=&\left(\frac{n+1}{n}\right)\tf(\mbox{untapered}),
\eeqa
where it must be kept in mind that $n>0$.

We shall focus on the case $n=1$, for which
the formation time is doubled over the
untapered
value
given in \S \ref{sec:power}. We note that the exponential
tapering factor used by \citet{mye98} can be approximated by
the $n=1$ case for early times. In this case, one can solve
for $t(m)$, obtaining
\beq
t(m)=\frac{2\tfo(m^{1-j}/\mf^{\jf-j})}{1+[1-(m/\mf)^{1-j}]^{1/2}}
~~~(n=1).
\label{eq:tm}
\eeq
Evaluation of $1-t/t_f$ allows one to express the accretion rate in
terms of only the masses. Defining
\beq
\delta_{n1}=\left\{ \begin{array} {r@{\quad\quad} l}0 & {\rm untapered},\, n=0,\\
	1 & \mbox{tapered},\, n=1, \end{array}  \right. 
\eeq
we can express the accretion rate for both the tapered and untapered cases as
\beq
\dot m=\mdo\left(\frac{m}{\mf}\right)^j\mf^\jf\left[1-\delta_{n1}\left(\frac{m}{\mf}\right)^{1-j}
\right]^{1/2}.
\label{eq:mdt}
\eeq
The star-formation time for these two cases is
\beq
t_f=(1+\delta_{n1})\tf(\mbox{untapered}).
\label{eq:tfa}
\eeq

\subsection{Mean Protostellar Mass}
\label{sec:mass}

Having described the accretion histories of the protostars, it
is possible to infer the mean protostellar mass,
\beq
\avg{m}_p=\int_0^\mup m \psi_p(m)d\ln m.
\label{eq:meanmass}
\eeq
Note that this is distinct from $\avg{\mf}$, the final protostellar mass
averaged over the IMF.
However, because the PMF is given in terms of an integral (eq. \ref{eq:pmf2}), it is more
straightforward to evaluate $\avg{m}_p$ using the bivariate PMF,
\beqa
\avg{m}_p &=&  \int_\ml^\mup d\ln\mf \int_0^\mf d\ln m \;m \ppt(m,\mf) \label{meanmass2} \\
&=&  \int_\ml^\mup d\ln\mf  \frac{m_f^{j-j_f} \psi(m_f) }{\mdo \avg{t_f}} \times\nonumber \\
& &~~~ \int_0^\mf d\ln m\, \frac{m^{2-j}}{\left[1-\delta_{n1}\left(m/\mf\right)^{1-j}\right]^{1/2}}. 
\eeqa

\begin{figure}
\epsscale{1.20}
\plotone{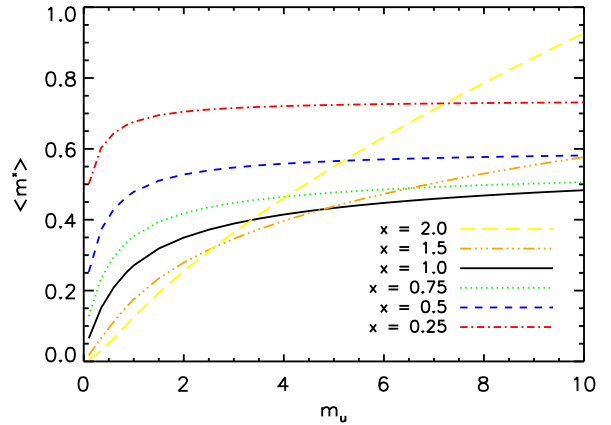}
\caption{ \label{mpower} 
Average values of powers of the protostellar mass,
$\avg{m^x}$, weighted by the \citet{cha05} IMF
for different powers of $x$ as a function of the upper mass limit in the cluster, $\mup$.} 
\end{figure}

The average star-formation time that enters this expression is
\beq
\avg{t_f}=\avg{\mf^{1-\jf}}(1+\delta_{n1})\tfo,
\eeq
from equations (\ref{eq:tf}) and (\ref{eq:tfa}).
For a non-tapered accretion history ($\delta_{n1}=0$), 
the expression for the average
protostellar mass reduces to
\beq
\avg{m}_p= \left( \frac{1-j }{2-j}\right) \frac{\avg{\mf^{2-j_f}}}{\avg{\mf^{1-j}}} 
\label{meanmassexp},
\label{eq:mp}
\eeq
which must be evaluated numerically.
In Figure \ref{mpower}, we have plotted $\avg{m^x}$ for 
the \citet{cha05} IMF (see \S \ref{sec:chab}) for
different exponents as a function of maximum mass in the cluster to facilitate the evaluation of equation (\ref{meanmassexp}).

\section{Results for the Steady-State PMF}

\subsection{Analytic Results for a Salpeter IMF}

In order to see how the different accretion histories affect the PMF, we
first consider a Salpeter IMF \citep{sal55},
\beq
\psal(\mf)\simeq 1.35\left(\frac{\ml}{\mf}\right)^{1.35}~~~~(\ml\leq\mf\leq\mup),
\eeq
where we have neglected the factor $(\ml/\mup)^{1.35}$ compared to unity
in the normalization. For the case in which $\mup$ is very large,
the average mass $\avg{\mf}=(1.35/0.35)\ml$; thus, to get an
average mass of $0.5 M_\odot$, for example, requires $\ml=0.13 M_\odot$.
To compare with results from the truncated Chabrier IMF below, we
also consider the case in which $\mup=3 M_\odot$ and 
$\avg{\mf}=0.4 M_\odot$; this requires $\ml=0.16 M_\odot$.

Since we are looking for simple analytic results in this subsection, we
focus on the case of untapered accretion.
The average star-formation time is then (eq. \ref{eq:tf})
\beq
\avg{t_f}=\frac{1.35\ml^{1-\jf}\phi(\jf)}{0.35+\jf}\;\tfo,
\eeq
where
\beq
\phi(\jf)=\left[1-\left(\frac{\ml}{\mup}\right)^{0.35+\jf}\right].
\eeq
For Isothermal Sphere accretion, $\avg{t_f}$ is just $\avg{\mf}\tfo$, or about half the time to form
a one solar mass star if the mean mass is about $0.5 M_\odot$.

With the aid of equation (\ref{eq:mdot}), the PMF in equation (\ref{eq:pmf2}) becomes
\beqa
\psi_p(m)&=& \frac{m}{\avg{t_f}}\int_\mfl^\mup\frac{\psal(\mf)}{\dot m}\; d\ln\mf,\\
&=&\frac{1.35 \ml^{1.35} m^{1-j}}{\mdo\avg{t_f}}\int_\mfl^\mup \frac {d\ln\mf}{\mf^\alpha},\\
&=& \frac{(1-j)(0.35+\jf)\ml^{0.35+\jf}m^{1-j}}{\alpha\phi(\jf)\mfl^\alpha}\;f(\mfl),~~~~~
\eeqa
where 
\beq
\alpha\equiv 1.35+\jf-j
\eeq
and
\beq
f(m)\equiv 1-\left(\frac{m}{\mup}\right)^\alpha.
\eeq
For masses large enough to be included in the IMF ($m>\ml$), we have $\mfl=m$,
so that
\beq
\psi_p(m)\propto m^{-(0.35+\jf)}f(m).
\label{eq:psisal}
\eeq

There are several points to note about this result
for the PMF. First, since it is weighted by
the accretion time, it is much flatter in the mass range $\ml\leq m\ll\mup$
for Isothermal Sphere accretion ($\psi\propto m^{-0.35}$) than for
Turbulent Core accretion
($\psi\propto m^{-1.1}$) or Competitive Accretion ($\psi\propto m^{-1.35}$);
indeed, 
since all stars have the same formation time in the latter model, it
just mimics the original Salpeter IMF. 
The Turbulent Core and Competitive Accretion
models are shifted to lower masses compared to the Isothermal Sphere accretion model
since they have longer accretion times at low mass, as shown in equation (\ref{eq:tacc1}).

Second, the PMF becomes depleted as $m\rightarrow\mup$, since $f(m)\rightarrow 0$
there. This occurs because protostars with final masses close to $\mup$ spend
most of their lives at lower masses as they grow by accretion; only a small
number of protostars are actually in the final stages of growth.

Third, the PMF is independent of the overall rate of star formation.
This was clear from our general expression for the PMF (eq. \ref{eq:pmf2}), in
which $\psi_p\propto \tacc/\avg{t_f}$ depends on the ratio of two star-formation times.

The final point to note is that the coefficient in $\psi_p(m)$ can be small,
particularly as $j\rightarrow 1$; thus, most of the protostars can
be at low masses. Evaluation of the mean protostellar mass
using equation (\ref{meanmassexp})
shows that it is much less for the Turbulent Core and Competitive
Accretion models than for the Isothermal Sphere model. For example, consider
the case in which 
$\ml=0.16M_\odot$ and $\mup=3M_\odot$; we find that the average
protostellar mass is $(0.38,\,0.16,\, 0.10)\, M_\odot$ for
Isothermal Sphere accretion, the Turbulent Core model and Competitive Accretion,
respectively.

\subsection{Results for the Chabrier IMF}
\label{sec:chab}

The Salpeter IMF has the benefit of being easily integrable,
but it is inaccurate at low masses.
In this section we derive the mean mass for each accretion model using the \citet{cha05} IMF:
\begin{eqnarray}
\psi_C &=& \psi_1 \exp-\left[\frac{(\log m-\log 0.2)^2}{2 \times 0.55^2}\right] \nonumber \\
&&~~~~~~~~~~~~~~~~~(m_l \le m \le 1 M_\odot), \\
\psi_C &=& \psi_2 m^{-1.35} ~~~~~(m  \ge  1 M_\odot) .
\end{eqnarray}	
The Chabrier IMF is log-normal below 1 $\msun$ and Salpeter above. 
The constants are determined by continuity and by enforcing: $\int_{m_l}^{m_u} \psi(m)\; d\; \mbox{ln}\ m = 1$. In this section, we adopt fiducial values of 
$m_{\ell} \sim 0.033~\msun$\footnotemark ~and $m_u = 3.0~\msun$, which yield $\psi_1 \simeq 0.35$ and $\psi_2 \simeq 0.16 = \psi_C (1)$. With these coefficients, $\psi_C$ gives an average stellar mass 
(including brown dwarfs) of
$\avg{m_*} \simeq 0.4~\msun$

\footnotetext{This limit is derived assuming a  $0.1 ~\msun$ minimum observable envelope mass, which converts its gas into a star with an efficiency factor of $\epsilon ~1/3$.
Reducing $m_\ell$ to the theoretically expected value of about $0.004 M_\odot$ 
\citep{whi07} has a small
effect on the mean mass (at most a factor 1.1 for the Competitive Accretion model)
and the median mass (a factor 1.25 for the same model); the ratio of the mean to median
thus drops by no more than a factor 1.15.}

\begin{figure}
\epsscale{1.20}
\plotone{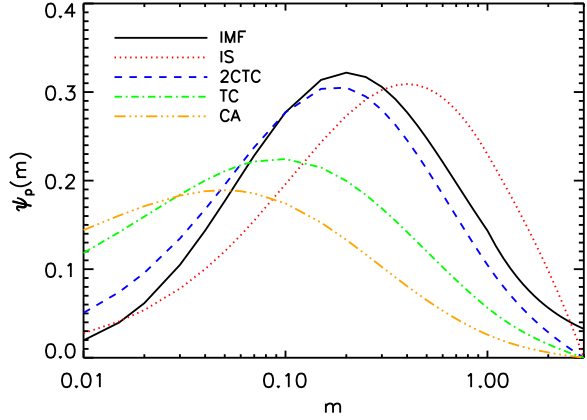}
\caption{ \label{pmf} The PMF for the four models with untapered accretion where $m_u = 3~\msun$. The Chabrier IMF is given by the solid line. For the Two-Component Turbulent Core model, we adopt $\calr_{\dot m}= 3.6$.}
\end{figure}

The PMF derived in equation (\ref{eq:pmf1}) can be expressed in terms of the Chabrier IMF, $j$, $j_f$, and $m_u$. 
The Two-Component Turbulent Core model also depends upon the ratio
$\calr_{\dot m}=\dot m_{\rm TC}/ \dot m_{\rm S}$.
For the numerical examples presented in this paper, 
we set the star-formation times for the Isothermal Sphere and Turbulent Core models
to be equal, which the data in Table 2 show corresponds to $\calr_{\dot m}=3.6$.
$\crdm$ exceeds unity because it is evaluated for $1 M_\odot$ stars; 
for this value of $\crdm$, the ratio
of the accretion rates tor the two models is close to unity for protostars of typical mass.
Estimates of the star-formation time have typically been based on the Class 0 lifetime, where most of the mass accretion was assumed to take place. However, observations by \citet{enoch08} indicate that the protostellar luminosities in the Class 0 and Class I phases are not substantially different, suggesting that significant accretion continues through
much of the Class I phase. 
For untapered accretion, the PMF (eq. \ref{eq:pmf1}) for the IS, TC and CA models is
\begin{equation}
\psi_p(m) = (1-j)m^{1-j} \frac{\int_\mfl^\mup \psi_C(\mf) \mf^{j-j_f}\; d\ln\mf} 
{\int_\ml^\mup\psi_C(\mf)\mf^{1-\jf}\; d\ln\mf},
\end{equation}
where we used equations (\ref{eq:mdot}) and (\ref{eq:tf}) to evaluate the the acceleration time
($\tacc\propto 1/\dot m$) and the mean formation time that enter the PMF.
For tapered accretion, a factor $[1-(m/\mf)^{1-j}]^{1/2}$ must be included in
the integrand in the numerator (see eq. \ref{eq:mdt}).
We plot the PMF for the four models in Figure \ref{pmf} assuming that $m_u = 3 \msun$. 
The Isothermal Sphere PMF peaks significantly to the right of the
other models.  As a result, it predicts a higher fraction of
relatively massive protostars in the distribution than any of the
other models, including the Chabrier IMF. This is a direct consequence
of the weighting by the star-formation time, in which more massive
stars have the longest accretion times in the Isothermal Sphere
case. The Competitive Accretion and Turbulent Core PMFs contain fewer
massive protostars than 
the Isothermal Sphere PMF because protostars in those models accrete
more rapidly as they approach their final mass, thereby reducing the
time they spend in the PMF.

\begin{figure}
\epsscale{1.20}
\plotone{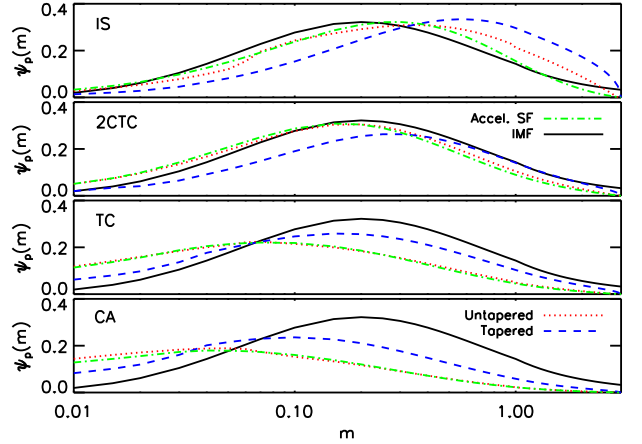}
\caption{ \label{allpmf} The PMF for the four models with untapered accretion(dotted),  tapered accretion (dashed), and untapered accretion with accelerating star formation (dot-dashed),  where $m_u = 3~\msun$. The Chabrier IMF is given by the solid line. For the Two-Component Turbulent Core model, we adopt $\calr_{\dot m}= 3.6$. For the accelerating star formation models we use $\tau = 1$ Myr.}
\end{figure}

The mean protostellar mass, $\avg{m}_p$, is given by equation (\ref{eq:mp}).
With the Chabrier IMF,
the Isothermal Sphere, Turbulent Core, and Competitive Accretion models give $\avg{m}_p = 0.47, 0.16, 0.09 \msun$, respectively, assuming that $m_u = 3~\msun$. 
The Isothermal Sphere mean value is
larger than that derived in the previous section using the Salpeter IMF with $m_l = 0.16 \msun$
since the Chabrier IMF turns over at $0.2 \msun$, so that the lower masses contribute less weight to the mean than in the case of the Salpeter IMF. The ordering of the values of the mean mass for these three models follows from the ordering of the values of the accretion
time (eq. \ref{eq:tacc1}), since $\tacc$ is the weighting factor that enters the PMF
(eq. \ref{eq:pmf2}) and it is largest at small masses for the Competitive Accretion
model and smallest for such masses for the Isothermal Sphere model.

\begin{figure}
\epsscale{1.20}
\plotone{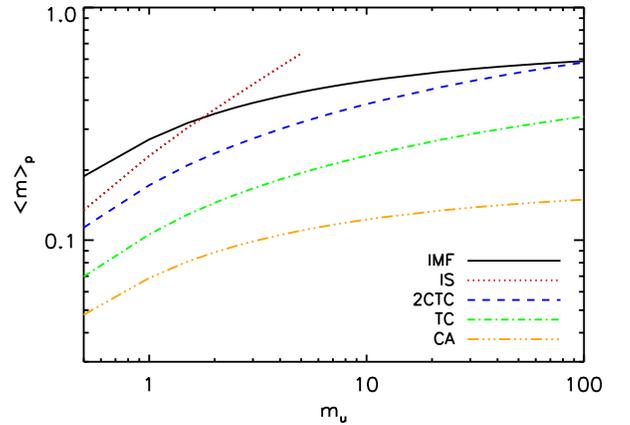}
\caption{ \label{meanmassplot} The mean mass (eq. \ref{meanmassexp}) of the PMF, $\avg{m}_p$,  for the four models as a function of  $m_u$
in the untapered, non-accelerating case.
The Chabrier IMF mean is shown by the solid line. For the Two-Component Turbulent Core model we adopt $\calr_{\dot m}= 3.6$. 
We restrict the Isothermal Sphere curve to $\mup\leq 5~\msun$.}
\end{figure}

Figure \ref{meanmassplot} shows $\avg{m}_p$  as a function of the maximum cluster mass. 
The Isothermal Sphere model has a significantly higher mean protostellar mass
than the stellar IMF as a result of the long formation times of more massive stars. In the Competitive Accretion and Turbulent Core models, 
the accretion rate accelerates with time so that protostars spend a larger fraction
of their lifetime at low masses than in the Isothermal Sphere case, in which
the accretion rate is independent of mass.
Note that we assume that the Isothermal Sphere model is suitable only for stars below $5 ~\msun$. When the accretion rate is modulated by turbulence, as in the Two-Component Turbulent Core model, the mean mass rises less steeply  than for a pure isothermal sphere as a function of cluster mass upper limit.

\begin{figure}
\epsscale{1.20}
\plotone{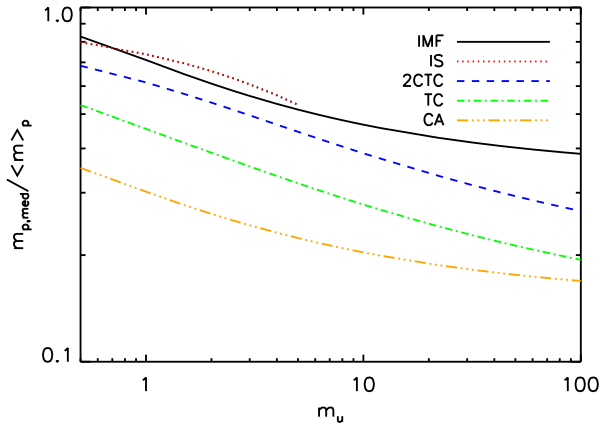}
\caption{ \label{medmassplot} 
The ratio of the median to mean protostellar mass of the PMF
for the four models as a function of $m_u$ with non-accelerating untapered accretion.
We adopt $\calr_{\dot m} = 3.6$
model. The value for the Chabrier IMF is 
indicated by the solid line. Since the Isothermal Sphere model is primarily for low-mass stars, we restrict it 
to $\mup\leq 5~\msun$.
}
\end{figure}

Figure \ref{medmassplot} shows the
ratio of the median to the mean 
protostellar mass as a function of the maximum stellar mass in the
cluster, $m_u$, for each of the accretion models. 
This ratio decreases with $m_u$ since
the mean mass increases with $m_u$ whereas
the median mass is relatively insensitive to it.
Since larger clusters can sample the rarer, more massive stars in the IMF,
we expect
large clusters to have small ratios of the
median to mean protostellar masses, particularly for the
Competitive Accretion and Turbulent Core models.
Furthermore, 
since the accretion  luminosity is proportional to the mass and the accretion rate, this
graph suggests that the Protostellar Luminosity Functions will be quite different for
the different models, although we expect
the differences to be less for tapered accretion (Paper II).

\begin{figure}
\epsscale{1.20}
\plotone{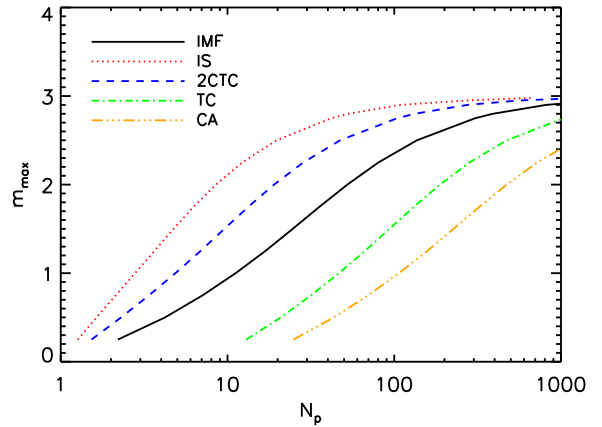}
\caption{ \label{npplot} The maximum protostellar mass, $\mmax$ as a function of the number of protostars, $\caln_p$ 
in the untapered, non-accelerating case.
The maximum stellar mass is $m_u = 3~\msun$. We adopt  $\calr_{\dot m}= 3.6$ 
for the Two-Component Turbulent Core model.
}
\end{figure}

The maximum expected protostellar mass, $\mmax$, in a cluster of $\caln_p$ protostars
is given by
\beq
\frac{1}{\caln_p} = \int_{\mmax}^\mup d\ln\mf\psi_p(\mf) \,\\
\label{eq:mmax}
\eeq
The resulting values of $\mmax$ are portrayed in Figure \ref{npplot}. The plot illustrates that
the expected value of the maximum protostellar mass,
$\mmax$, for the Competitive Accretion and Turbulent Core models with $\caln_p\la 100$ stars
can be significantly less than the maximum stellar mass, $\mup$,
which is determined by the Class II stars in the cluster that obey a normal IMF.

\section{Accelerating Star Formation}
\label{sec:accel}

From an analysis of the pre-main-sequence stars in the Orion Nebula Cluster,
\citet{pal99} concluded that the star formation there has been accelerating.
\citet{pal00} extended this analysis to seven other star-forming regions
and found evidence for acceleration in all but one case. They attributed this
acceleration to contraction of the parent molecular cloud, which they surmised
was a quasi-static process. They inferred exponentiation times ranging
from 1.0~Myr for $\rho$~Oph and IC~348 to 3.3~Myr for NGC~2264.
The shortest of these acceleration times is not that much greater than
the typical star-formation time of 0.54~Myr found by \citet{eva09}.
Here we determine the effect of accelerated star formation on the PMF.

The evolution of the PMF in time is governed by a continuity equation.
To conform with standard practice, we define $n(m,\mf,t)dm d\mf$
as the number of protostars in the mass range $dm$ and final mass 
range $d\mf$ at
time $t$. Under the assumption that the stars are born
with an initial mass function $\psi(\mf)$, the continuity equation for $n$ is then
\beq
\ppbyp{n}{t}+\pbyp{m}( n\dot m) = \dot\caln_*(t)\delta(m)\;\frac{\psi(\mf)}{\mf},
\eeq
where $\dot\caln_*(t)$ is the rate at which stars are born at time $t$,
$\delta(x)$ is the delta function, and the factor $\mf^{-1}$ allows
from the conversion from $n\,d\mf$ to $\psi \,d\ln\mf$.
The Green's function for this problem, which we denote by
$G(t-t_0)$, is the solution for
$\dot\caln_*(t)=\delta(t-t_0)$. Writing the protostellar mass as an explicit
function of time, $m=\mu(t-t_0)$,
we have
\beq
G(t-t_0)=\delta[m-\mu(t-t_0)]\,\frac{\psi(\mf)}{\mf}\, H(t-t_0),
\eeq
where $H(t-t_0)$ is the Heaviside step function. The general solution
is then
\beqa
n&=&\int_{-\infty}^\infty G(t-t_0) \dot\caln_{*,0}  dt_0,\\
&=& \int_{-\infty}^t \delta[m-\mu(t-t_0)]\,\frac{\psi(\mf)}{\mf}\, \dot\caln_{*,0} dt_0.~~~~
\eeqa
Note that $t-t_0$ is the age of a star at time $t$ that was born at
time $t_0$. Let $t_m$ be the age of a star of mass $m$ and
final mass $\mf$. 
We can then rewrite the $\delta$-function as
\beq
\delta[m-\mu(t-t_0)]=\frac{\delta(t-t_0-t_m)}{\dot m},
\eeq
where $\dot m=d\mu/dt$ is the accretion rate. The solution is then
\beq
n(m,\mf,t)=\frac{\psi(\mf)\dot\caln_*(t-t_m)}{\mf\dot m}.
\eeq

To convert this to the bivariate PMF, note that
\beq
\ppt(m,\mf,t)=\frac{n(m,\mf,t) \,m\mf}{\int_\ml^\mup d\mf\int_0^\mf dm\, n(m,\mf,t)},
\eeq
so that
\beq
\ppt=\frac{\psi(\mf)\displaystyle\frac{m}{\dot m}\, \dot\caln_*(t-t_m)}{
\int_\ml^\mup d\ln\mf\psi(\mf)\int_0^{t_f} dt_m\,\dot\caln_*(t-t_m)}.
\label{eq:ppta}
\eeq
In a steady state [$\dot\caln_*(t-t_m)=$~const], this reduces to the result given in
equation (\ref{eq:ppt}). 
The protostellar mass function for an arbitrary star formation history
is then
obtained by inserting this result into equation (\ref{eq:pmf1}).

As a simple model of accelerating star formation, we assume an
exponentially increasing birthrate,
\beq
\dot\caln_*(t-t_m)=\dot\caln_{*,0}e^{(t-t_m)/\tau}, \label{eq:dndt}
\eeq
where 
$\dot\caln_{*,0}=\dot\caln_{*}(t=0)$ 
is the current birthrate. Substituting into equation (\ref{eq:ppta}),
we find
\beq
\ppt=\frac{\psi(\mf)\tacc\exp-(t_m/\tau)}{\tau\avg{1-\exp(-t_f/\tau)}}.
\eeq
The PMF is then
\beqa
\psi_p(m)&=&\frac{1}{\tau\avg{1-\exp(-t_f/\tau)}} \times \nonumber\\
& &\int_\mfl^\mup \psi(\mf)\tacc e^{-t_m/\tau} d\ln\mf,
\eeqa
which reduces to equation (\ref{eq:pmf1}) for $\tau\rightarrow\infty$.

The PMF depends on the time scale $t_m$,
the age of a protostar of mass $m$ and final mass $\mf$.
If the accretion is untapered, then
the protostellar mass grows according to 
equation (\ref{eq:mass}) and 
$t_m$ is given by
\beq
t_m=\tfo\,\frac{m^{1-j}}{\mf^{\jf-j}}.
\eeq
For the Two-Component Turbulent Core model, $t_m$ is given by
\beq
t_m={{2} \over{\dot m_S}} {\sqrt{\calr_{\dot m}^2 \mf^{1/2}m +1} -1 \over {\calr_{\dot m}^2\mf^{1/2}}}.
\eeq
The value of $t_m$ for the case of tapered accretion has been given in equation (\ref{eq:tm}).
Figure \ref{allpmf} shows $\psi_p(m)$ in the accelerating case with untapered accretion. 
We consider the case with $\tau = 1$ Myr, which is approximately twice the average star-formation time.
In comparison to the steady star formation (dotted lines), the PMF peaks are shifted towards lower masses for accelerating star formation except in the Competitive Accretion case, where the peak mass increases from $\sim 0.05 \msun$ to $\sim 0.08 \msun$.

\section{Comparison with Observations \label{sec:obs}}

\subsection{The Star Formation Timescale}
As discussed in the Introduction, it is not presently possible to directly measure 
the masses of the protostars in a star-forming region. The protostellar luminosity 
function is subject to direct observation and will be discussed in Paper II.
Nonetheless, it is possible to carry out an approximate comparison between observation
and theory by comparing
the average star-formation time observed
in low-mass star-forming regions with the theoretical values predicted in \S \ref{sec:accretion}.

First, we show that the average observed star-formation time as determined by
number counts is the IMF-averaged formation time, $\avg{t_f}$, not the
PMF-averaged formation time, $\avg{t_f}_p$. \citet{eva09} determined the
average star-formation time, $\avg{t_f}_{\rm obs}$, by comparing the number of
protostars with the number of Class II sources:
\beq
\frac{\avg{t_f}_{\rm obs}}{\avg{t_\ii}}=\frac{\caln_p}{\caln_{\rm II}} \label{eq:tobstII},
\label{eq:tftii}
\eeq
where $\caln_{\rm II}$ and $\avg{t_\ii}$ are the number and average lifetime of Class II sources.
Although nothing
is known at present about the mass dependence of the Class II lifetime, we
allow for the possibility that there is such a dependence by using the IMF-averaged value,
$\avg{t_\ii}$, in this equation.
The number of protostars is just the star-formation rate times the 
IMF-averaged lifetime (eq. \ref{eq:np}),
and similarly, the number of Class II sources is just $\caln_{\rm II}=\nds \avg{t_\ii}$. As
a result, we have
\beq
\frac{\avg{t_f}_{\rm obs}}{\avg{t_\ii}}=\frac{\nds\avg{t_f}}{\nds \avg{t_\ii}}=
\frac{\avg{t_f}}{\avg{t_\ii}},
\label{eq:tfobsii}
\eeq
so that
\beq
\avg{t_f}_{\rm  obs}=\avg{t_f}.
\eeq
Thus, the mean formation time inferred from the ratio of the number of
protostars to the number of Class II sources is equal to the
IMF-averaged value of the formation time.
By contrast, if there were a 
method of determining the formation time of each observed protostar,
the average of these times, $\avg{t_f}_p=\int d\ln m \psi_p(m) t_f$, would be
quite different. We generalize this expression to the case of accelerating star formation in the Appendix.

\begin{deluxetable*}{lcccc}
\tablewidth{0pt}
\tablecolumns{5}
\renewcommand*\arraystretch{1.5}
\tablecaption{Star-Forming Region Properties  \label{tablecloud}}
\tablehead{  
\colhead{Region} & \colhead{$T$ (K)\tablenotemark{a}} &
\colhead{$\Sigma$ (g cm$^{-2}$)\tablenotemark{b}} &\colhead{$\bar
  n_{\rm H}$ ($10^4$ cm$^{-3})$\tablenotemark{c}}
&\colhead{$\avg{t_{\rm 0,I}}$ (Myr)\tablenotemark{d} }}
\startdata
 \\[-1ex]
Perseus & 10-13  	& 0.06 &1.0 &0.72 \\ 
Serpens& 10-15	&  0.06 & 0.9 &0.56 \\
Ophiuchus & 12-20 & 0.08  &  1.4  &0.40 \\ [1ex] \hline \\ [-1ex]
Average 				&       & 0.07	&  1.1   &
0.56 
\enddata
\tablenotetext{a} {The citations for the temperatures in each of the
 regions are \citet{foster09}; \citet{enoch08}; \citet{andre07}}
\tablenotetext{b} {Average gas column of the clumps with $A_{\rm V}
 \ge 10$, where the clumps are assumed to be spherical. Masses and sizes are supplied
by \citet{eva07} and \citet{enoch07}. 
The gas with $A_{\rm V} \ge 10$ approximately corresponds to the minimum column density for star formation: $N({\rm H}_2)=
8\times 10^{21}$ cm$^{-2}$ \citep{onishi98}, where 
$A_{\rm V} = N({\rm H}_2)/ (0.94 \times 10^{21}$~cm\ee).
}
\tablenotetext{c} {The densities are calculated by identifying clumps
 of gas with $A_{\rm V}\ge
 10$. The average density of each region is calculated assuming
 spherical symmetry. The mass weighted average of the clumps is
 reported here.  Data are derived from \citet{eva07}
 and \citet{enoch07}.}
\tablenotetext{d} {Average extinction corrected (Class 0 + Class I) lifetimes \citep{eva09}}
\end{deluxetable*}

\citet{eva09} report an average
star-formation time, $\avg{t_f}$, of
0.54 Myr after correcting for extinction for five local star-forming
regions. This value is the sum of the estimated Class 0 and Class I
lifetimes and assumes a Class II lifetime of 2 Myr. 
\citet{eva09} find that the lifetimes vary significantly over their sample of clouds, where the lowest lifetimes correspond to the smallest clouds containing the fewest protostars.  The lifetimes have an uncertainty of order $\pm$ 0.1 Myr.
In our comparison, we focus on Perseus, Ophiuchus, and
Serpens, which are much more massive than Lupus and Cha II. They have a
combined average star-formation time of 0.56 Myr.

As discussed in the Introduction, by adopting the protostellar
lifetimes from \citet{eva09}, we implicitly assume a direct mapping
between Class 0, I and Stage 0, I. 
The core mass
estimates obtained for these regions by \citet{enoch09} 
support this assumption for both Perseus and Serpens since all 
but one Class I envelope exceed $0.1 \msun$
in Perseus and none in Serpens.
In contrast, \citet{enoch09} report that about half of the
$\rho$ Ophiuchus Class I sources have envelopes with masses 
$< 0.1 \msun$, 
so that they are not true protostars by our definition.
This suggests
that the value of $\avg{\tf}_{\rm obs}$ we adopt from \citet{eva09},
which includes the Class I sources with envelopes less than $0.1 M_\odot$,
overestimates the mean protostellar lifetime in that region. 
Since Ophiuchus already has the shortest lifetime of the three clusters we consider, 
it is possible that the rate of star formation in this cluster has recently slowed down.
Bearing in mind these caveats,
in our comparison
we adopt conservative error estimates that are comparable to the error
resulting from the inclusion of the small-envelope sources.

To compare with the observations, we use the more evolved Class II sources to estimate the cluster upper mass limit, $m_u$. Since the Class II sources are about three times more numerous than the Class 0 and Class I objects, their luminosities provide a reasonable expectation for the most massive star forming in the cluster.
\citet{eva09}  report  a maximum extinction corrected Class II
luminosity of $\sim 90 \lsun$, suggesting a maximum stellar mass of
$\sim 3 \msun$ \citep{pal99}. As a counterpoint, if we assume that the combined population of
Class II sources is drawn from the Chabrier IMF with maximum
stellar mass of 120 $\msun$ then a statistical argument suggests that 12 of the
Class II objects should have masses greater than $3~\msun$. Perseus,
containing 244 Class II sources, should have five such stars. 
This difference between the observed IMF and the expected one is partially due to selection, since the clusters were chosen to have only low-mass stars. It is also possible that the
conditions in these clouds work against the formation of massive stars.

We use the 
model definitions from Table \ref{table1} to infer
the physical parameters 
for each model given an average star-formation time of  $\avg{t_f} =
0.56$ Myr. We give the observed 
physical parameters values for
local molecular clouds in Table 3.
Each of the models constrains a different physical parameter (Table \ref{tablemod}). 
For the Competitive Accretion model, the relevant star-formation time, $\avg{t_f}$, is about equal to the freefall time of the entire clump, so that the lifetime depends only upon the average density.  
For the Two-Component Turbulent Core model, we 
assume that the thermal and turbulent contributions to the accretion are comparable
by setting 
$\crdm=3.6$ (see \S \ref{sec:chab}).
(The Turbulent Core 
model was developed for high-mass star formation, and the Turbulent Core contribution
grows in importance as the stellar mass increases, eq. \ref{eq:mdtctc}.)

To compare the models with accelerating star formation with observation, we first
use a least-squares approach to re-analyze the data assembled by \citet{pal00}, who
fit their data ``by eye." 
We find $\tau=0.9$ Myr for Ophiuchus, in good agreement with the \citet{pal00} value of $\tau =1$ Myr. At face value, this suggests that star formation is rapidly accelerating. However, the ages of stars included in the fit are greater than 0.5 Myr, so that one cannot use this trend to reliably infer information about more recent star formation activity. To emphasize this point, observations  by \citet{enoch09} find only 3 Class 0 sources, which, when compared to the relatively larger numbers of Class I sources, suggest that current star formation in Ophiuchus is likely {\it decelerating}. A fit of IC348, an active region of Perseus, gives $\tau=2.2$ Myr  rather than 1 Myr as reported by  \citet{pal00}.  Several other regions discussed by \citet{pal00} show a turnover in the number of stars at recent times, further highlighting the point that the star formation rate may not be monotonic and is not necessarily well represented by a single exponential.
Despite this caveat, we select $\tau =2$ Myr as a fiducial value.   For Perseus and Ophiuchus, we individually adopt $\tau= 2$ Myr and $\tau = 1$ Myr, respectively.
We adopt 2 Myr for the Class II lifetime, which is the same value \citet{eva09} use to estimate the observed protostellar lifetime. 
Note that it is possible that the Class II lifetime depends on stellar mass, but in
the absence of any relevant data we neglect this possibility here.

The results summarized in Tables 3-5 show the inferred physical conditions in
an average star-forming cloud
(like Serpens), in Perseus and in Ophiuchus for each model
for the cases of tapered and untapered accretion, accelerating and non-accelerating
star formation. The inferred temperatures for the Isothermal Sphere and
Two-Component Turbulent Core models are closer to the observed values when
tapering and/or acceleration is allowed for. 
Tapering
or allowing for acceleration gives a good fit for both the
Turbulent Core model and the Competitive Accretion model 
in Perseus.
In Serpens, non-accelerating star formation with untapered or tapered
accretion gives good
agreement for these models.
In Ophiuchus, both models do best for the untapered, non-accelerating
case. 

However,
as discussed in \S \ref{sec:accretion} above, the accretion rates for the various models
are uncertain
for several reaons: they do not allow for the effects of protostellar winds in
disrupting the protostellar cores, they do not allow for an initial infall velocity, and
they do not include the effects of magnetic fields.
As a result of these competing effects, the fiducial accretion rate coefficients in Table 2 
may be either higher
or lower than those we have adopted, thus impacting our estimates
of the physical parameters. 
If we assume that this uncertainty is a factor two in the accretion rate, then this
corresponds to a factor $2^{2/3}\simeq 1.6$ in the inferred temperature in the Isothermal Sphere model, a
factor $2^{4/3}\simeq 2.5$ in the inferred column density in the Turbulent Core model, and
a factor $2^2=4$ in the inferred density in the Competitive Accretion
model. 
It must also be borne in 
mind that there are uncertainties in the observed values of the physical parameters
describing the clouds,
particularly the lifetimes. 
With allowance for an overall factor of two uncertainty in the
accretion rate relative to the actual values, 
we summarize the consistency between the
models and observed regions in Table \ref{tablesummary}.
We find that the Isothermal Sphere model is consistent with
the data for both Perseus and Ophiuchus for all but the non-accelerating, untapered model.
One might question this consistency, given that the temperature is known to much
better than the factor 1.6 we are allowing; however, we are allowing for uncertainty
in the coefficient of $T^{3/2}$ in equation (\ref{eq:mds}), not for errors in the temperature.
The Two-Component
Turbulent Core model is consistent with the data for 
all  but the non-accelerating, untapered model;
it would have the same level of consistency with the data as the IS model if
$\rmd$ were lowered from 3.6 to 1.7. 
(Based on eq. \ref{eq:tftc}, this corresponds to increasing the
value of $\mf$ for which $t_f$ is closer to the TC case than to the IS case
from about $0.5\, M_\odot$ to about $1.2\, M_\odot$.)
The Turbulent Core and Competitive Accretion models 
are consistent with the Perseus data for all but the non-accelerating, tapered model, but they are consistent with the Serpens and
Ophiuchus data only for the non-accelerating models.

We note that the free-fall time corresponding to the mean density in
the regions we have considered is $\tff\simeq 0.4$~Myr, which is comparable to the
mean star formation times (which is consistent with Competitive Accretion), but
short compared to the acceleration time. Conceptually, the Competitive Accretion model
involves rapid acceleration, but numerical simulations are needed to determine
whether the observed relatively long acceleration times are consistent with the model.

\begin{figure}
\epsscale{1.20}
\plotone{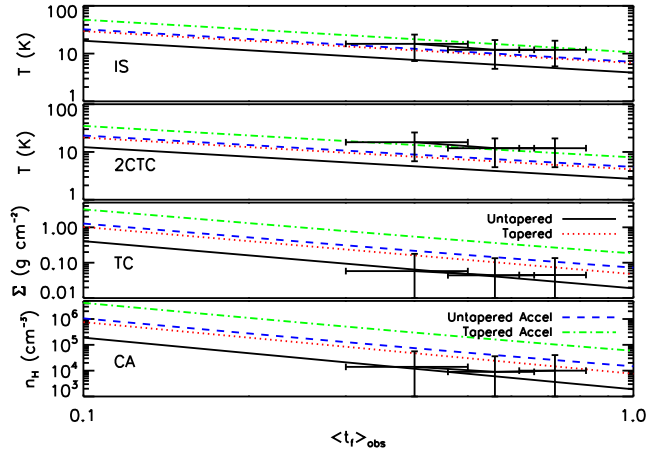}
\caption{ \label{tfobsvparam} The physical parameter for each model
 versus the observed star-formation time, $\avg{t_f}_{\rm obs}$. We
 assume $n=1$, $\tau = 1$ Myr, and $\avg{t_\ii} = 2$ Myr in the
 tapered and accelerating cases. 
Ophiuchus (left), Serpens (middle), and Perseus (right) are overlaid
with horizontal error bars for the uncertainty in the
measurements and vertical error bars for the uncertainty due to
the model accretion rates.
}
\end{figure}

Since a number of the
models predict reasonable physical parameters, uncertainty in
the mean temperatures, densities, columns, and lifetimes 
contributes to our inability to determine which models are consistent with observation.
For comparison in the context of these uncertainties, we plot the
physical parameters of each of the model as a function of $\avg{t_{\rm
   obs}}$ in Figure \ref{tfobsvparam}. This plot also illustrates that in all cases the
inferred temperature, column density, and density
    increase with decreasing star-formation times.

\subsection{Discussion}

In this section, we address various sources of uncertainty inherent in the calculations.
As described in \S 3.1, the instantaneous accretion rate may be larger than estimated at early times and may fluctuate strongly on short timescales. We have also neglected 
possible temporal
variations in the core star formation efficiency--i.e., the fraction of the mass in the core that
actually accretes onto the star--in modeling the protostellar accretion rate.
Finally, the values of $n$, $\avg{t_\ii}$ and $\tau$  have some intrinsic uncertainty and may vary between clouds.  We discuss these three parameters below.

Tapering of the accretion is very plausible, since
we expect accretion to decrease as the infall phase ends and the core mass is depleted. 
Measurements of the protostellar luminosity appear inconsistent with constant or increasing accretion through the Class I phase,
a topic we shall address in Paper II. However, there are no direct estimates of $n$, and the exact tapering function is poorly constrained by observation.  Unfortunately, we find that we cannot definitively constrain $n$ on the basis of the observed star-formation time due to variation between clouds and uncertainties in both the physical parameters and the accretion 
models, as discussed above.

\citet{eva09} estimate an uncertainty of $\pm 1$ Myr for the Class II
lifetime, a variation of 50\%.   As shown by equation
(\ref{eq:tftii}),
the observed star-formation time varies directly as $\avg{t_\ii}$ in the non-accelerating
case. The accretion rate then varies as $1/\avg{t_\ii}$, so that
$T^{3/2}\propto 1/\avg{t_{\rm II}}$ in the Isothermal Sphere case, 
$\Sigma^{3/4}\propto 1/\avg{t_{\rm II}}$ in the Turbulent Core case,
and $\nh^{1/2}\propto 1/\avg{t_\ii}$ in the Competitive Accretion
case. Thus, 
a longer Class II lifetime would decrease the inferred temperature, column density and density accordingly. 
Our analysis  in \S 6.1 suggests that a longer Class II phase would
worsen agreement of the Isothermal Sphere and Two-Component Turbulent Core models
with the observations, since the implied temperatures
are already somewhat low in most cases. Conversely, a shorter Class II lifetime
would improve agreement with observation for 
many of these models, with the exception of the tapered, accelerating Isothermal
Sphere model for Ophiuchus.

As discussed in \S 5, \citet{pal00} derived $\tau$ values between $\sim
1-3$ Myr for a collection of eight clouds. They performed the fits by
eye so that their estimates of the e-folding time are not precise. 
We used a least-squares fit to get improved values for the acceleration times, but it
must be borne in mind that the actual star formation histories may be much more complex
than the simple exponential form that we have used to fit them.

\begin{deluxetable*}{llllll}
\tablewidth{0pt}
\tablecolumns{6}
\tablecaption{Model Properties for $\avg{\tf}_{\rm obs}=0.56$ Myr \label{tablemod}}
\tablehead{ 
\multicolumn{2}{c}{} & \multicolumn{2}{c}{Non-Accelerating}&
\multicolumn{2}{c}{Accelerating\tablenotemark{a}} \\ 
\colhead{Model} & \colhead{Parameter}& \colhead{Untapered} &
\colhead{Tapered\tablenotemark{b}}
&\colhead{Untapered } & \colhead{Tapered\tablenotemark{b}}}
\startdata
Isothermal Sphere  & 	$T$(K)	&  5.9 & 9.3 & 7.8& 12 \\
2C Turbulent Core \tablenotemark{c} & $T$(K) & 3.9	&  6.1 &  5.4& 8.6 \\
Turbulent Core (TC)  & $\Sigma$ (g cm$^{-2}$)	& 0.041 & 0.10 & 0.083& 0.21 \\
Competitive Accretion & $\bar n_{\rm H}$ ($10^4$ cm$^{-3}$)& 0.60 & 2.4& 1.8& 7.1  
\enddata
\tablenotetext{a} {The fiducial $\avg{t_f}$ is the same as
  $\avg{\tf}_{\rm obs}$ for Serpens, 
  so the data for
  Serpens should be compared with
  these values.}
\tablenotetext{b} {$\tau = 2$ Myr; $\avg{t_\ii} = 2$ Myr}
\tablenotetext{c} {$n=1$}
\tablenotetext{d} {We fix $\crdm=3.6$.}
\end{deluxetable*}

\begin{deluxetable*}{lllllll}
\tablewidth{0pt}
\tablecolumns{7}
\tablecaption{Model Properties for $\avg{\tf}_{\rm obs}=0.72$ Myr: Perseus  \label{tablemodper}}
\tablehead{ 
\multicolumn{3}{c}{} & \multicolumn{2}{c}{Non-Accelerating} &
\multicolumn{2}{c}{Accelerating \tablenotemark{a}}\\ 
\colhead{Model} & \colhead{Parameter}& \colhead{Observed Value }
&\colhead{Untapered } & \colhead{Tapered\tablenotemark{b}}
&\colhead{Untapered }& \colhead{Tapered\tablenotemark{b}}}
\startdata
Isothermal Sphere  & 	$T$ (K)	& 10-13	                  &  5.0  & 7.9 & 6.6&  10  \\
2C Turbulent Core\tablenotemark{c} & $T$ (K) &10-13	  & 3.3	  &
5.2 &  4.6 & 7.3  \\
Turbulent Core (TC) & $\Sigma$ (g cm$^{-2}$)&	0.06	  & 0.029 & 0.074 & 0.061 & 0.15  \\
Competitive Accretion &$\bar n_{\rm H}$ ($10^4$ cm$^{-3}$)&1.0 	  & 0.37  & 1.5 & 1.1&  4.5  \enddata
\tablenotetext{a} {$\tau = 2$ Myr, the value for IC 348; $\avg{t_\ii} =
 2$ Myr}
\tablenotetext{b} {$n=1$}
\tablenotetext{c} {We fix $\crdm=3.6$.}
\end{deluxetable*}

\begin{deluxetable*}{lllllll}
\tablewidth{0pt}
\tablecolumns{7}
\tablecaption{Model Properties for $\avg{\tf}_{\rm obs}=0.40$ Myr: Ophiuchus  \label{tablemodoph}}
\tablehead{  
\multicolumn{3}{c}{} & \multicolumn{2}{c}{Non-Accelerating} &
\multicolumn{2}{c}{Accelerating \tablenotemark{a}}\\ 
\colhead{Model} & \colhead{Parameter}& \colhead{Observed
 Value}&\colhead{Untapered} & \colhead{Tapered\tablenotemark{b}}
&\colhead{Untapered }& \colhead{Tapered\tablenotemark{b}}}
\startdata
Isothermal Sphere  & 	$T$ (K)	&12-20 	                    & 7.4  & 12 & 13 & 20\\
2C Turbulent Core \tablenotemark{c} & $T$ (K) &12-20 	    & 4.8 &  7.7 &  8.7 & 14\\
\enddata
Turbulent Core (TC)  & $\Sigma$ (g cm$^{-2}$) &0.08 	    &0.064 & 0.16 & 0.22& 0.55 \\
Competitive Accretion & $n_{\rm H}$ ($10^4$ cm$^{-3}$) &1.4   &1.2  &4.7  & 7.4 &  30 
\tablenotetext{a} {$\tau = 1$ Myr; $\avg{t_\ii} = 2$ Myr}
\tablenotetext{b} {$n=1$}
\tablenotetext{c} {We fix $\crdm=3.6$.}
\end{deluxetable*}
\bigskip

\begin{deluxetable*}{lllllllllllll}
\tablecaption{Model Summary  \label{tablesummary}}

\tablehead{  
\colhead{} &  \multicolumn{4}{c}{Ophiuchus} & \multicolumn{4}{c}{Serpens} & \multicolumn{4}{c}{Perseus} \\ 
\colhead{Model} & \colhead{UN \tablenotemark{a}}&
\colhead{TN}&\colhead{UA} & \colhead{TA}&  \colhead{UN}&
\colhead{TN}&\colhead{UA} & \colhead{TA}& \colhead{UN}&
\colhead{TN}&\colhead{UA} & \colhead{TA} }

\startdata
Isothermal Sphere& \checkmark   &\checkmark  &\checkmark  &\checkmark   & \checkmark       &\checkmark &\checkmark &\checkmark  &o       &\checkmark &\checkmark & \checkmark    \\

2C Turbulent Core& o   &\checkmark  &\checkmark  &\checkmark  & o
&\checkmark  &\checkmark  &\checkmark &  o     &\checkmark &\checkmark
&\checkmark    \\

Turbulent Core (TC)& \checkmark   &\checkmark &o &o & \checkmark        &\checkmark
&o &o & \checkmark      &\checkmark &\checkmark & o     \\

Competitive Accretion& \checkmark   &\checkmark & o& o&  \checkmark
&\checkmark &o &o & \checkmark      &\checkmark &\checkmark & o    \\ 
\enddata
\tablenotetext{a} {UN = Untapered Non-accelerating,TN = Tapered
 Non-accelerating, UA = Untapered Accelerating, TA = Tapered Accelerating }

\end{deluxetable*}

\section{CONCLUSIONS}

We have analyzed the Protostellar Mass Function (PMF), which is the
Present-Day Mass Function (PDMF) 
of a cluster of protostars. The PMF builds on
the IMF, which measures the final mass distribution of a cluster of stars after
completion of the formation process. The PMF depends on the 
mass dependence of the formation time of the stars, $t_f$, and therefore can
in principle provide an observational diagnostic of theories of star formation.
Direct observation of the PMF must await further observational and theoretical
advances, since it is not currently possible to
infer the mass for protostars because
they are generally extremely obscured and 
the luminosity is dominated by accretion. Nonetheless, it is possible to
compare theoretical models
with measurements of the mean lifetime of protostars in different molecular
clouds as we have done here. 
Furthermore, the PMF is the basis for the Protostellar Luminosity Function,
which is directly accessible to observation (Paper II).

We have made two key assumptions in inferring the PMF.
First, since no stars above 3 $\msun$
are seen in the low-mass clusters we have analyzed
\citep{eva09, enoch09}, we have
assumed that none of the smaller number of protostars in these clusters will
grow to more than $3 \msun$; we therefore
imposed an upper cutoff of $3 M_\odot$ on the IMF in our analysis.
Given the number of Class II sources, which have very nearly reached their final masses,
we note that 12 Class II sources with $m>3\msun$
would be expected in these clusters based on a \citet{cha05} IMF.
These clusters were selected on the basis of their proximity so as to allow 
a thorough study of the Class 0, I, and II sources, so it is unlikely that
this anomalous IMF is due to a selection effect.
The second key assumption we have made is that the
accretion rate is a simple function of only the current protostellar mass, 
the final mass and the time. We thus do not allow for variations in
the accretion rate due to 
a brief high-accretion Larson-Penston phase or to temporal fluctuations
in the accretion rate
(although insofar as such fluctuations are random
and there is a statistically large sample of protostars, they should not significantly
affect the PMF).

We consider four accretion rate histories:
the classical Isothermal Sphere accretion \citep{shu77}, the
Turbulent Core model \citep{mck02,mck03}, a blend of the two
(Two-Component Turbulent Core, 2C Turbulent Core), and 
an analytic approximation for the
Competitive Accretion model \citep{bon97,bon01a}. There are substantial
uncertainties in the accretion rates for each model: In all cases, one
must allow for the effect of protostellar outflows, which can reduce the 
accretion rate by a factor of a few \citep{mat00}. For the first three, there
is a countervailing correction needed to allow for an initial infall velocity.
Our approximation for the Competitive Accretion model captures many
of its essential features, but since the model itself
is based primarily on numerical simulations, there is
no fully analytic form for it.  In comparing the models with observation,
we assume that
the star formation is steady or accelerating; 
since the Competitive Accretion model has been developed
for the evolution of individual star clusters, the comparison with observation
is valid for this model only
if a number of clusters are sampled, either because a forming cluster is comprised
of a number of sub-clusters or because data from different clusters are averaged together.

 The mean protostellar mass
(Fig. \ref{meanmassplot}) and the ratio of the median mass to the mean
mass (Fig. \ref{medmassplot}) depend sensitively on the accretion history.
The Turbulent Core and Competitive Accretion Models have accretion
rates that increase with mass and therefore with time ($\dot m\propto 
m^j$, with $j=\frac 12,\, \frac 23$ for the two models respectively). As a result,
protostars of a given final mass, $\mf$,
spend a smaller fraction of their lives at high mass than in
the Isothermal Sphere model. Furthermore, these two models have accretion
rates that increase with $\mf$, so that it takes less time to form a high-mass
star than in the Isothermal Sphere model. Both effects are stronger
for the Competive Accretion model. As a result, the mean protostellar mass
increases systematically from the Competitive Accretion model to the Turbulent Core
model to the Two-Component Turbulent Core model to the Isothermal Sphere model.
The ratio of the median to mean protostellar mass follows the same ordering,
and the same effect shows up in the plots of the PMF in Figure \ref{pmf}.

A common feature of 
all the accretion models is that the accretion rate
remains constant or (usually) increases until the time at which the protostar reaches its
final mass, when it abruptly ceases. In reality, as pointed out
by \citet{mye98}, the accretion will turn off gradually. To allow for this,
we have inserted  a factor $[1-(t/t_f)^n]$ into the accretion rates; we refer to
this as tapered accretion. In practice, we focused on the case $n=1$, which
gives an accretion time $t_f$ twice as long as would be expected in
the absence of tapering. This has the effect of increasing the temperature,
column density or density of the model needed to match a given observed
formation time. For example, in the Isothermal Sphere model, the accretion
rate is proportional to $T^{3/2}$, so the temperature needed to match the
observations of a given formation time is $2^{2/3}$ times greater for a
tapered model than for an untapered one. Not only is tapering
physically plausible, it also generally results in models
that are in better agreement with observation. 
As shown in Figure \ref{allpmf}, tapering moves the peak of the PMF to higher masses
since stars spend a larger fraction of their lives at high mass when the accretion
slows down at the end of the accretion process.

The rate of star formation should accelerate in time in a contracting
gas cloud, and \citet{pal00} found direct evidence for such acceleration
in a number of nearby star-forming clusters. We generalized our analysis
of the PMF to the case in which the star formation rate is time dependent
in \S \ref{sec:accel}. For simplicity, we have assumed that the acceleration applied
only to the rate at which stars formed, not to the accretion rates of individual
stars. This is a reasonably good approximation for the cases we analyzed, which
have star-formation times that are significantly smaller than the time scale
for acceleration. Moreover, the approximation is even better for
for the Isothermal Sphere model,
since the accretion rate depends on the temperature and radiative losses maintain
an approximately constant temperature.  On the other hand,  the time scale
for acceleration is often comparable to or less than the mean lifetime of
Class II sources, so the ratio of the number of protostars to the number
of Class II sources is larger than in the non-accelerating case.
As a result, as shown in the Appendix, the ``observed" star-formation time,
which is given by equation (\ref{eq:tftii}), exceeds the actual
star-formation time. 
We find that acceleration does not have a substantial effect on the PMF. Rather, its
primary effect is to
reduce the inferred time scale for the formation of individual stars, thereby
increasing 
the inferred temperature, column density or mean density, depending
on the accretion model.

In the absence of any direct information on protostellar masses, we were 
able to carry out only a very crude comparison with observation: Using the
observed star-formation time scales in two different clusters, we
computed the implied temperature (Isothermal Sphere model), surface density 
(Turbulent Core model),
and mean density (Competitive Accretion model), and then compared with 
the observed values of these parameters. We found that the tapered accretion
and accelerating star formation models were somewhat better than
untapered, non-acceleration models, but we could
not draw any firm conclusions due to uncertainties 
in both the observations and in the models,
which have accretion
rates that are probably uncertain by a factor of 2.
In addition, the molecular clouds have an unknown internal structure
and the IMF can have significant statistical and perhaps physical
fluctuations from one cloud to another.
In Paper II we shall show that the Protostellar Luminosity Function is a
more powerful diagnostic for inferring the accretion mechanism.

\acknowledgments
We thank Steve Stahler for pointing out his previous work on this problem,
Neal Evans, Shu-Ichiro Inusuka and Zhi-Yun Li for useful comments,
and Melissa Enoch for clarifying and supplying observational values for the data comparison.
This research has
been supported by the NSF through grants AST-0606831 (CFM \& SSRO),
AST-0908553 (CFM), and AST-0901055 (SSRO).

\appendix

\section{Observed Star Formation Time for Accelerating Star Formation}

For a star formation rate that varies exponentially in time, the number of protostars
is given by 
\beqa
\caln_p(t=0)&=&\int_\ml^\mup dm_f\int_0^\mf 
dm\,
n(m,\mf,t=0)  \\
&=& \dot\caln_{*,0} \tau\int d\ln\mf \psi(\mf)  \left(1-e^{-\tf
/\tau}
\right)\\
&=& \dot\caln_{*,0} \tau\avg{1-e^{-t_f/\tau}}.
\label{eq:accnp}
\eeqa
In the time-dependent case, 
the number of Class II sources in the
mass range $d\mf$ is the number formed in the time interval $\tf<t_0<\tf+t_\ii$,
\beq
d\caln_\ii(\mf)=\psi(\mf)d\ln\mf \int_{\tf}^{\tf+t_\ii} \nds(t-t_0)\, dt.
\eeq
For an exponentially increasing star formation rate (eq. \ref{eq:dndt}),
the total number of Class II sources is then
\beq
\caln_\ii=\ndso\tau\left\langle e^{-\tf/\tau}\left(1-e^{-t_\ii/\tau}\right)\right\rangle,
\eeq
where the average is over the IMF.
With the aid of equation (\ref{eq:accnp}), equation (\ref{eq:tftii}) then
implies that the mean observed star-formation time is
\beq
\avg{\tf}_\obs=\frac{\avg{t_\ii}\left\langle 1-e^{-t_f/\tau}\right\rangle}
{\left\langle e^{-\tf/\tau}\left(1-e^{-t_\ii/\tau}\right)\right\rangle}.
\label{eq:tfobsacc}
\eeq
In the limit of steady star formation ($\tau\rightarrow \infty$), this approaches
the actual value of the average star-formation time, $\avg{\tf}$. 
For $\tf\ll\tau$, which is true for most of the models we have considered,
the observed star-formation time is related to the actual value by
\beq
\avg{\tf}_\obs\simeq \left[\frac{\avg{t_\ii}}{\tau\left(1-e^{-\avg{t_\ii}/\tau}\right)}\right]\avg{\tf} \label{eq:tfobsacc2},
\eeq
where we have made the approximation $\avg{1-\exp(-t_\ii/\tau)}
\simeq 1-\exp(-\avg{t_\ii}/\tau).$
The observed 
formation time, $\avg{tf_\obs}$, thus
exceeds the actual value, $\avg{t_f}$, since the number of Class II sources is 
suppressed.
In the text we evaluate equation (\ref{eq:tfobsacc}) for $t_\ii=2$~Myr for each of
the accretion cases considered.

\end{document}